\documentclass[sigconf]{acmart}
\usepackage[ruled,vlined,linesnumbered]{algorithm2e}
\usepackage{multirow}
\usepackage{booktabs}
\usepackage{enumitem} 
\usepackage{comment}
\usepackage{url}
\usepackage{caption}
\usepackage{tabularx}
\usepackage{subcaption}
\usepackage{xcolor}
\usepackage{footmisc}
\usepackage{array}
\usepackage{amsmath}
\usepackage{float}
\colorlet{RED}{red}
\colorlet{OLIVE}{olive}
\DeclareMathOperator*{\argmax}{arg\,max}
\DeclareMathOperator*{\argmin}{arg\,min}

\makeatletter
\newcommand{\removelatexerror}{\let\@latex@error\@gobble}
\makeatother
\AtBeginDocument{%
  \providecommand\BibTeX{{%
    \normalfont B\kern-0.5em{\scshape i\kern-0.25em b}\kern-0.8em\TeX}}}
\makeatletter
\newcommand\footnoteref[1]{\protected@xdef\@thefnmark{\ref{#1}}\@footnotemark}
\makeatother
\copyrightyear{2024}
\acmYear{2024}
\setcopyright{rightsretained}
\acmConference[WWW '24 Companion]{Companion Proceedings of the ACM Web
Conference 2024}{May 13--17, 2024}{Singapore, Singapore}
\acmBooktitle{Companion Proceedings of the ACM Web Conference 2024 (WWW
'24 Companion), May 13--17, 2024, Singapore,
Singapore}\acmDOI{10.1145/3589335.3648318}
\acmISBN{979-8-4007-0172-6/24/05}

\newcommand{\model}{PP-GLAM}
\newcommand{\fullmodel}{Plug and Play Graph LAnguage Model}
\begin{document}

\title{An Interpretable Ensemble of Graph and Language Models for Improving Search Relevance in E-Commerce}

\settopmatter{authorsperrow=4}
\author{Nurendra Choudhary}
\orcid{0000-0002-4471-8968}
\affiliation{%
  \institution{Virginia Tech, Amazon}
  \city{Arlington}
  \state{VA}
  \country{USA}
}
\email{nurendra@vt.edu}

\author{Edward W Huang}
\affiliation{%
  \institution{Amazon}
  \city{Palo Alto}
  \state{CA}
  \country{USA}
  }
\email{ewhuang@amazon.com}

\author{Karthik Subbian}
\affiliation{%
  \institution{Amazon}
  \city{Palo Alto}
  \state{CA}
  \country{USA}}
\email{ksubbian@amazon.com}

\author{Chandan K. Reddy}
\orcid{0000-0002-4471-8968}
\affiliation{%
  \institution{Virginia Tech, Amazon}
  \city{Arlington}
  \state{VA}
  \country{USA}
}
\email{reddy@cs.vt.edu}

\renewcommand{\shortauthors}{Choudhary, et al.}
\begin{abstract}

{The problem of search relevance in the E-commerce domain is a challenging one since it involves understanding the intent of a user's short nuanced query and matching it with the appropriate products in the catalog. This problem has traditionally been} addressed using language models (LMs) and graph neural networks (GNNs) to capture semantic and inter-product behavior signals, respectively. However, the rapid development of new architectures has created a gap between research and the practical adoption of these techniques. Evaluating the generalizability of these models for deployment requires extensive experimentation on complex, real-world datasets, which can be non-trivial and expensive. Furthermore, such models often operate on latent space representations that are incomprehensible to humans, making it difficult to evaluate and compare the effectiveness of different models. This lack of interpretability hinders the development and adoption of new techniques in the field.
To bridge this gap, we propose {\fullmodel} ({\model}), an explainable ensemble of plug and play models. Our approach uses a modular framework with uniform data processing pipelines. It employs additive explanation metrics to independently decide whether to include (i) language model candidates, (ii) GNN model candidates, and (iii) inter-product behavioral signals. For the task of search relevance, we show that {\model} outperforms several state-of-the-art baselines as well as a proprietary model on real-world multilingual, multi-regional e-commerce datasets. To promote better model comprehensibility and adoption, we also provide an analysis of the explainability and computational complexity of our model. We also provide the public codebase and provide a deployment strategy for practical implementation.
\end{abstract}



\maketitle

\section{Introduction}
\label{sec:intro}
\begin{figure}
      \centering
      \includegraphics[width=.75\linewidth]{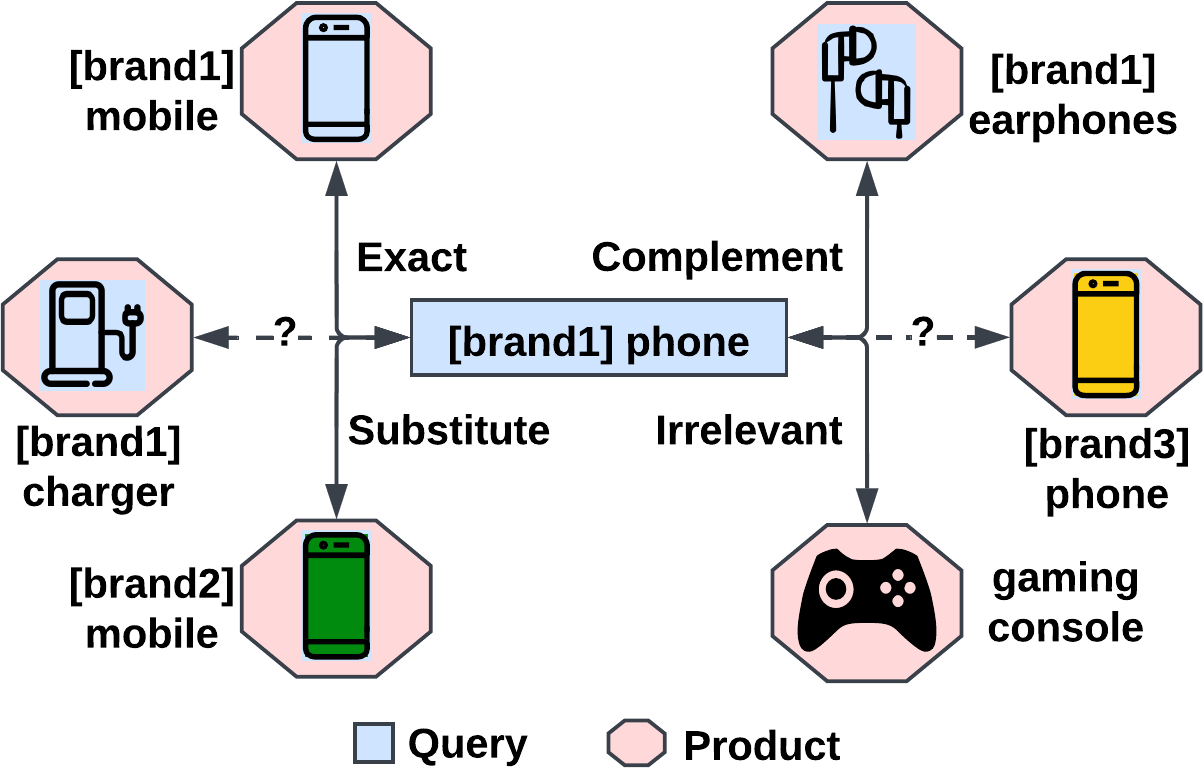}
      \vspace{-1em}
      \caption{An illustration of the search relevance problem. Using the semantic information of queries and products along with the behavioral relationships between them, the goal is to classify the degree of relevance for unseen query-product pairs (labeled with question marks) as exact, substitute, complement, or irrelevant.}
      \label{fig:esci_example}
      \vspace{-1em}
  \end{figure}

  The rapid advancements in language modeling and graph neural network research have created challenges for traditional industry frameworks looking to adopt the latest developments. Adopting a new model requires extensive experimentation to determine its generalizability on practical datasets, but the ever-changing nature of these datasets makes it difficult to compare old models with new ones. As a result, all models and datasets must be carefully considered before determining if a new model can be adopted in practice. This presents a significant challenge for industry frameworks seeking to incorporate the latest advancements in the field. In this study, we focus on the specific issue of search relevance\footnote{We chose search relevance due to the practical impact of the problem and the availability of public data. Our proposed methods are extensible to other problems.} in the e-commerce industry, where the goal is to classify the relationship between a query and a product as one of exact, substitute, complement, or irrelevant (ESCI), as shown in Figure \ref{fig:esci_example}.
  Human annotation for the purpose of training classification models in the domain of e-commerce search (ESCI) can be a resource-intensive task. The datasets that are typically used for training these models tend to be significantly smaller than those used for training relevance models, which often rely on anonymous, aggregated customer shopping behavior data. This is particularly true for labeled data that is available for training on a per-region basis, which tends to be even smaller in size. However, e-commerce datasets do contain additional information that may be useful for model training, such as query-item graphs that capture interactions between queries and items, and co-purchase behavior across products in a catalog. At first glance, search relevance may appear to be a straightforward multi-classification task. However, the short text nature of queries and products \cite{choudhary2022self} as well as the influence behavioral signals have on their relationship make it a non-trivial problem. 
  
  Initial work in this field \cite{subhi2021survey,nigam2019semantic} relied on the semantic features of queries and products, but subsequent research has highlighted the importance of incorporating behavioral signals \cite{choudhary2022graph} such as click-through data and purchase data in learning representations. However, these methods combine semantic and graph information in a latent space, which results in a lack of interpretability. Furthermore, they are standalone models that must be individually trained and evaluated for their specific scenario. This approach faces three major challenges: (i) the dynamic nature of real-world data requires frequent and costly updates to the deployment model, often requiring the retraining of all previous candidates to maintain comparability, (ii) certain decisions, such as which semantic/behavioral signals to include, cannot be automated, and (iii) model operations in a latent space result in limited interpretability of the results.
  
  To overcome these challenges, \textit{we propose {\fullmodel} ({\model}), a modular ensemble of LMs and GNNs} that relies on additive explanation values to automatically decide on the inclusion of semantic and behavioral signals. Our model uses an ensemble of LMs \cite{he2021deberta} to capture semantic information from the query and product. While previous approaches \cite{nigam2019semantic,choudhary2023complex} also utilize LMs, they generally default to a multilingual framework due to the multilingual nature of e-commerce datasets. However, LMs specifically developed on resource-rich languages such as English tend to outperform their multilingual counterparts \cite{khanuja2021mergedistill}. Thus, it is advantageous to combine the benefits of language-specific models with multilingual ones. Consequently, in {\model}, we encode the query-product semantic information using multiple language-specific and multilingual LM models, and rely on a Gradient-Boosted Decision Tree (GBDT) for effective model selection. However, a majority of the language models only support a fixed number of tokens, which are not sufficient for handling long product descriptions. While recent works such as Longformers \cite{zaheer2020big} are able to circumvent this issue, they are computationally more expensive and add to both inference time and cost. Thus, in our model, we denoise the product description to a fixed token size using certain importance measures \cite{rose2010automatic}. More importantly, we also leverage several query-product signals such as clicks, purchases, add-to-carts (or adds), impressions, and consumes. These relations vary both in their density and correlation with target variables (e.g., clicks are more common than purchases, but purchases are more correlated to a query-product match). This density-correlation trade-off of relations must be thoroughly studied for every task and deployment iteration, which adds to the deployment time and effort. To alleviate this problem, we automate these decisions by considering all the potential homogeneous (single-relation) and heterogeneous (multi-relation) GNNs in a GBDT ensemble, subsequently eliminating the worst-performing models using their ranking over explainable SHAP values \cite{lundberg2017a}. Unlike MLP \cite{pal1992multilayer} and Attention models \cite{vaswani2017attention}, which aggregate features in a latent space, our GBDT based ensemble can be made explainable using additive SHAP values. The additivity allows us to independently decide the importance of individual LMs and {relational GNNs} on the target variable and {plug-and-play them to improve model performance}. Subsequently, automating the decision of including the best-performing models can be done based on the computational constraints. We conduct extensive experiments on large-scale e-commerce data to demonstrate that our ensemble framework significantly outperforms the state-of-the-art baselines in the e-commerce tasks of search relevance and irrelevant detection, respectively. Furthermore, we conduct several studies to analyze the computational complexity of our model. We also demonstrate our model's interpretability through reports of the target variable's dependence on the models and relations. Lastly, we provide a deployment strategy to apply our modular framework in a practical environment. To summarize, our main contributions are as follows:
  \begin{enumerate}[leftmargin=*]
      \item In this paper, we present {\model}, a modular ensemble of LMs and {relational GNNs} that utilizes GBDTs for flexible model selection in a practical environment.
      \item We demonstrate the effectiveness of {\model} on search relevance and irrelevant detection compared to state-of-the-art {public and proprietary} baselines on a multi-regional e-commerce dataset.
      \item We show the benefits of utilizing ensembles as an interpretable strategy to aggregate semantic and behavioral signals and efficiently select the most impactful models.
      \item We {provide a detailed strategy} to deploy our modular framework in a practical environment with dynamic data sources.
  \end{enumerate}
  The rest of the paper is organized as follows: Section \ref{sec:related} discusses the relevant background and Section \ref{sec:model} describes our proposed model architecture. Section \ref{sec:experimental} demonstrates our experimental results and subsequent analysis. Section \ref{sec:discussion} details our deployment strategy and Section \ref{sec:conclusion} concludes our work.

\section{Related Work}
\label{sec:related}
In this section, we discuss previous work related to our problem in the areas of language models, graph neural networks, and graph-enhanced language models.

\textbf{Language Models (LMs).} Transformer-based LMs are a type of neural network architecture that has revolutionized the field of natural language processing (NLP) in recent years. These models have achieved state-of-the-art performance on a wide range of NLP tasks such as machine translation \cite{bahdanau2015neural}, language generation \cite{brown2020language}, and search \cite{devlin-etal-2019-bert}. The development of transformer-based LMs originated from the introduction of word embedding models, such as Word2Vec \cite{mikolov2013distributed} and GloVe \cite{pennington2014glove}, which represent words as continuous vectors in a high-dimensional space to capture the semantics and relationships between words. These embeddings have been used as inputs to deep learning models, such as recurrent neural networks (RNNs) \cite{sutskever2014sequence} and convolutional neural networks (CNNs) \cite{shen2014learning}. However, such methods are limited in their ability to capture word sense and contextual semantics. Transformer-based models \cite{vaswani2017attention} were introduced in conjunction with RNNs and CNNs to capture the contextual dependencies between words in a sequence. However, further research \cite{vaswani2017attention} demonstrated their standalone effectiveness and led to the rise of task-tunable pre-trained LMs such as BERT \cite{devlin-etal-2019-bert} and XLNET \cite{yang2019xlnet}. Recent developments in LMs have focused on {multilingual extensions \cite{conneau2019cross} and increasing the size and complexity of the models. Given this rapid pace of advancement, it is important that industry frameworks can quickly adopt and deploy the latest LMs. Thus, we design {\model} as a flexible modular ensemble.

\textbf{Graph Neural Networks (GNNs).} GNNs are a special type of neural network architecture, designed to operate on graph-structured data. Early research in graph processing predominantly relied on matrix factorization \cite{ou2016asymmetric,nonlinear2000sam} and random walk-based \cite{grover2016node2vec,narayanan2017graph2vec,tang2015line} approaches. {Matrix factorization is a technique that utilizes linear algebra techniques \cite{klema1980the, Ortigueira1991eigen} to simplify the representation of the relationships between nodes in a graph by decomposing the adjacency matrix into a lower-dimensional, latent space}. Random walks use Markov chain models to simulate transitions between nodes and learn their representations. These techniques demonstrate the effectiveness of vector space models in graph representation learning. However, they are limited in their ability to capture node neighborhood relations. To address these limitations, GNNs were adopted to effectively aggregate node neighborhood features, typically for node classification \cite{scarselli2008graph}, link prediction \cite{kipf2017semi}, and graph classification \cite{nguyen2022universal}. These techniques include Graph Convolutions (GCNs) \cite{kipf2017semi}, Graph Attention (GAT) \cite{velickovic2018graph}, and GraphSage \cite{hamilton2017inductive}, which respectively use convolution filters, attention mechanisms, and graph sampling to learn node and link representations for downstream tasks. Current research in the area focuses on applying GNNs to heterogeneous graphs \cite{wang2019heterogeneous}, temporal graphs \cite{rossi2020temporal}, and knowledge graphs \cite{bordes2013translating}. {Given the short text nature of queries in the task of search relevance, we need to utilize behavioral graph signals to better understand users' intent. However, with the diversity of behavioral signals available in e-commerce datasets, it is difficult to select the most significant candidates for a downstream task. Therefore, our model utilizes SHAP values to select the set of best-performing signals and corresponding GNNs in a computationally constrained deployment environment.}

\textbf{Graph-Enhanced Language Models (GELM).} Recently, researchers have focused on integrating graph signals and text features in unified GELM models. Initial studies, such as TextGNN and TextGCN, used semantic features to initialize GNN models for downstream tasks such as node classification and link prediction. However, a primary limitation of this approach is its limited ability to learn task-specific 
semantic embeddings. To overcome this issue, Graphormers \cite{ying2021do} combined manual graph features, such as spatial encoding and centrality measures, with a text-based Transformer architecture to capture hybrid features. This allows the model to effectively handle both graph and text data. However, calculating the required graph features is computationally expensive and requires frequent model re-training with dynamic datasets. To avoid this, the (proprietary) SALAM model \cite{choudhary2022graph} uses an attention mechanism to combine graph and text features in a scalable industry setting. Furthermore, GreaseLM \cite{zhang2022greaselm} employs intermediate MLP units after each layer to combine graph (GNN layer) and text (LM layer) features. However, aggregations in such models occur in a latent space, and are hence not interpretable for decision-making. Thus, \textit{we developed {\model} for industry settings as a modular combination of LMs and {relational GNNs} that use interpretable SHAP values to decide on the model components.}
\begin{figure*}
    \centering
    \vspace{-1em}
    \includegraphics[width=.9\linewidth]{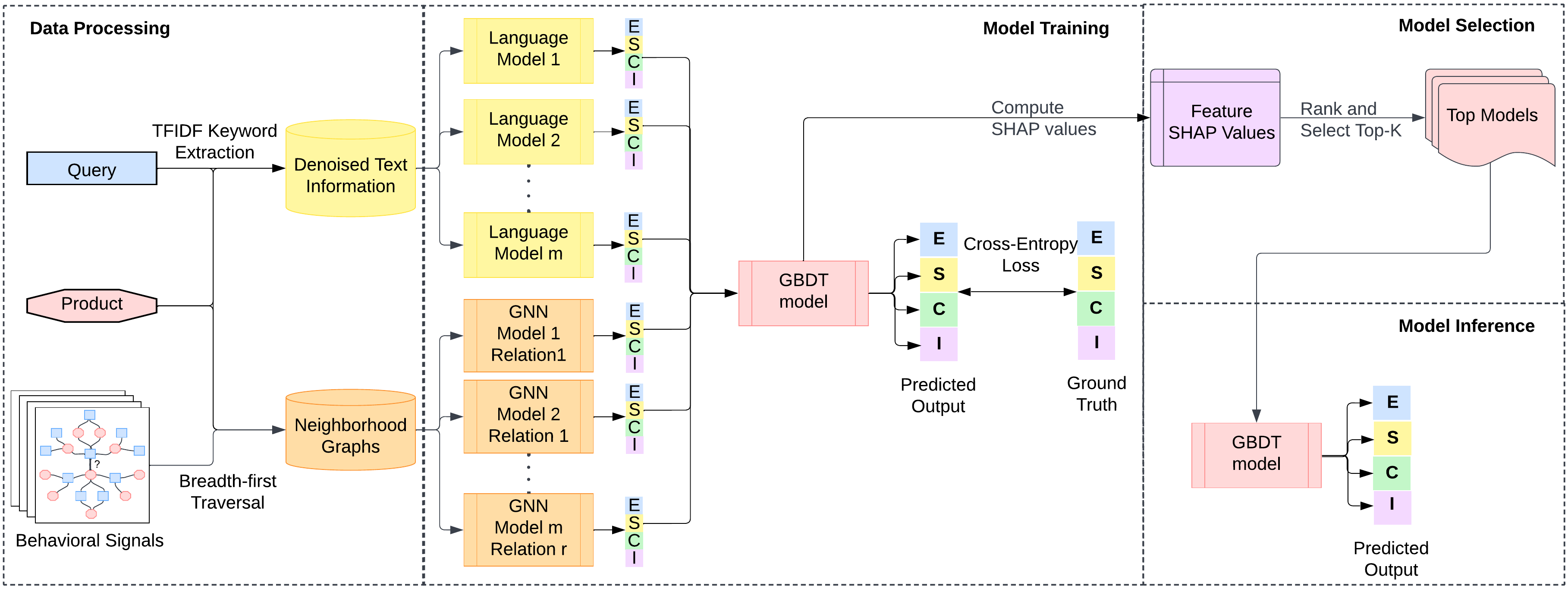}
    \vspace{-1em}
    \caption{An overview of the {\model} architecture. The model contains four modules: (i) Data Processing handles the pre-computation of graph neighborhoods and the de-noising of the product information, (ii) Model Training uses the training samples with query-product pairs and corresponding ground-truth labels to learn the parameters of the GBDT-based ensemble containing different language models and graph neural networks (GNNs), (iii) Model Selection module utilizes interpretable SHAP values to eliminate low-performing models based on the constraints of the inference setup, and (iv) Model Inference loads the selected set of models into memory for inference on new batches.}
    \vspace{-1em}
    \label{fig:ensemble_model}
\end{figure*}
\section{Proposed Framework}
\label{sec:model}
This section discusses the problem statement of the search relevance task, describes the different components of our model, and explains its training and inference pipeline.

\subsection{Problem Statement}
Let the set of query-product pairs be $(Q,P)$ and the corresponding query-product graphs of clicks, impression, adds, purchases, consumes, and a combined heterogeneous version (any) be denoted by their adjacency matrix $\mathcal{G}^{\psi}:Q\times P$, where signal $\psi\in\{$clicks, impressions, adds, purchases, consumes, any$\}$. Note that, ``any'' signal implies the presence of any of the mentioned relations. Each element ${G}^{\psi}_{qp}$ indicates the rate of signal $\psi$ between query $q \in Q$ and product $p \in P$ (e.g., the number of product clicks for a particular query). The objective of search relevance is to learn a model $P_\theta$ with parameters $\theta$ that classifies query-product pairs into the following relevance classes: exact (E), substitute (S), complement (C), and irrelevant (I). The model is formulated as:
\begin{align}
    \hat{y} = \argmax_{y'=\{E,S,C,I\}}P_\theta(y'|x,\theta);\quad\theta=\argmax_{\theta'} P_{\theta'}(\hat{y}=y|x,\theta')
\end{align}
{where $\theta'$ is a sample parameter set from the parameter search space,} $y, \hat{y} \in \{E,S,C,I\}$ are the ground-truth value and the model output from $P_\theta$ for an input $x=\left(q_i,p_j,G^{\psi}\right)$, respectively.
\subsection{Model Elements}
This section describes the different elements of our model and the ensemble of language and graph features for better generalization over diverse e-commerce datasets. Our model pipeline starts with the input query $x=(q,p)$ with which the corresponding subgraph $\mathcal{G}^{\psi}(q,p)$ is constructed using the Graph Extraction module. Subsequently, the textual information and subgraph are encoded using multiple language models $\{LM_{i}(x)\}_{i=1}^{|LM|}$ and {relational GNNs} $\{GNN_{j}(\mathcal{G}^{\psi}(q,p))\}_{j=1}^{|GNN|}$ to produce the corresponding ESCI labels $\hat{Y}=\left\{\hat{y}^{LM}_{1},\hat{y}^{LM}_{2},...,\hat{y}^{LM}_{|LM|},\hat{y}^{GNN}_{1},\hat{y}^{GNN}_{2},...,\hat{y}^{GNN}_{|GNN|}\right\}$. These label outputs $\hat{Y}$ and certain manual features (such as language information) are finally ensembled together in a GBDT model to produce the final {\model} output. 

\textbf{Graph Extraction.} In this module, we extract the local neighborhood subgraph pertaining to the input query-product pair $x = (q,p)$. In our problem, we use the signals of $\psi\in\{$clicks, impressions, adds, purchases, consumes, any$\}$, with a corresponding adjacency matrix $\mathcal{G}^{\psi}$ for each signal. To construct the local $k$-hop neighborhood subgraph $\mathcal{G}^{\psi}(q,p)$, we perform a breadth-first traversal around both $q$ and $p$ until a maximum depth of $k$. We observe that this process is independent for each $(q,p)$ pair, and hence, we parallelize it over the CPU cores and store the subgraphs as hash maps for constant time retrieval in the training and inference pipelines.

\textbf{De-noising Product Information.} Typical product data consists of long text descriptions that are not supported by generic language models. Hence, we utilize a TF-IDF \cite{rose2010automatic} pre-processor to rank the terms' importance in the description and eliminate words beyond the language model's supported capacity $\gamma$. This is formalized for a text sequence of tokens $\{t_i\}_{i=1}^{|p|} \in p$ as:
\begin{equation}
    TFIDF(t_i,p) = TF(t_i,p)\times\log\left(\frac{|P|}{DF(t_i,P)}\right)
\end{equation}
where $TF(t_i,p)$ is the number of times $t_i$ occurs in $p$ and $DF(t_i,P)$ is the number of $p\in P$ that contains $t_i$. Letting $TFIDF(t_\gamma,p)$ be the $\gamma$-th largest value in the sorted set of $\{TFIDF(t_i,p)\}_{i=1}^{|p|}$, then the filtered set $p' = \{t_i\in p | TFIDF(t_i,p)\geq TFIDF(t_\gamma,p)\}$. 

The query and product information is then merged with separator tokens as $t=q\|[SEP]\|p'$, processed through the cross-encoder tokenizers \cite{song2021fast}, and stored as hash maps for efficient retrieval in the model pipelines.

\textbf{Language Models.} E-commerce datasets are generally diverse since they typically span multiple geographical regions with multiple languages. Hence, in our model, we utilize an ensemble $\{LM_{i}\}_{i=1}^{|LM|}$ of language-specific LMs that work well for resource-rich languages (English), including DeBERTa \cite{he2021deberta}, COCOLM \cite{yu2022cocolm}, BigBird \cite{zaheer2020big}, and a multilingual LM - M-DeBERTa \cite{he2021debertav3}\footnote{\label{note:lm-gnn}Note that we only listed the LM and GNN models used in our experiments. The framework, however, supports most generic LM and GNN models.}.
The tokenized input sequence $t$ is encoded using the LM encoders $LM_i \in LM$ into its encoding $e\in \mathbb{R}^d$ with a softmax layer $\phi_d^l:\mathbb{R}^d\rightarrow\mathbb{R}^l$ to obtain the probability over $l$ output labels. The encoder is formalized as
\begin{align}
    \hat{y}^{LM}_i(t) &= \phi_d^l\left(e_i(t)\right),\quad\text{where }e_i(t) = LM_i(t) \label{eq:lm}\\
    \hat{Y}_{LM}(t) &= \left\{\hat{y}^{LM}_i(t)\middle| i \in [1,|LM|]\right\},\quad\text{where }\hat{y}^{LM}_i(t)\in\mathbb{R}^l \label{eq:lm_agg}
\end{align}

\textbf{Graph Models.} To aggregate the structure signals, we use an ensemble of $|\psi|$ behavior signals over the GNN networks of GCN, GAT, and GraphSage\footnoteref{note:lm-gnn}. For an input query-product graph with $v$ nodes and adjacency matrix $\mathcal{G}^{\psi}(q,p)\in\mathbb{R}^{v\times v}$, {let the feature matrix be $h_0\in\mathbb{R}^{v\times d}$ and $(i_q,i_p)$ be the indices of query and product node in the feature matrix, then $K$-layer GNN models aggregate the node neighborhood information as:}
\begin{align}
    h_{k+1} &= \sigma\left(D^{\zeta-1}\mathcal{G}^{\psi}(q,p)D^{\zeta}h_kW_k\right),\quad\text{where }\zeta\in[0,1]\\
    \hat{y}_i^{GNN}(h_0) &= \phi_d^l(h_{k+1}[i_q]\|h_{k+1}[i_p]),~~\text{where }\hat{y}^{GNN}_i(h_{0})\in\mathbb{R}^l \label{eq:gnn}\\
    \hat{Y}_{GNN}(h_{0}) &= \left\{\hat{y}^{GNN}_i(h_{0})\middle| i \in [1,|GNN|]\right\}\label{eq:gnn_agg}
\end{align}
where $\zeta$ is the factor of spectral filter dependent on the GNN model, $D \in \mathbb{R}^{v\times v}$ is the diagonal degree matrix and $\|$ is the concatenation operator. Note that we solve the search relevance as a link prediction task in GNNs and hence only use $(i_q,i_p)$ in the final softmax layer to compute the probabilities over $l$ output labels.

\textbf{Interpretable Ensemble.} To aggregate the features from different models in an interpretable manner, we utilize Gradient Boosting Decision Trees (GBDTs) \cite{ke2017lightgbm}. GBDTs are a type of ensemble machine learning algorithm that combines the predictions of multiple decision trees to improve the accuracy and stability of the model. The algorithm works by adding decision trees to the model sequentially, with each tree correcting the errors made by the previous tree. The trees are trained using gradient descent optimization, which minimizes the error between the predicted output and the ground truth. The final prediction is obtained by averaging the predictions of all trees in the ensemble. For an input text $t$, graph $h_0$, and associated region information $f$, the output of a GBDT model is formulated as:
\begin{equation}
    \hat{y} = \frac{1}{T}\sum_{t=1}^T o_t(\hat{Y}),\quad\text{where }\hat{Y}=\hat{Y}_{LM}(t)\|\hat{Y}_{GNN}(h_{0})\|f \label{eq:gbdt}
\end{equation}

where $\hat{y}$ is the final prediction, $T$ is the number of trees in the ensemble, and $o_t(\hat{Y})$ is the prediction of the $t$-th tree. In our task, we optimize the GBDT's model parameters $\theta$ by minimizing the cross-entropy loss for $l$-labels $Y=\{y_1,y_2,...,y_l\}$ as follows:
\begin{align}
    L(y,\hat{y}) &= -\sum_{y \in Y}\left(ylog(\hat{y})+(1-y)log(1-\hat{y})\right) \label{eq:loss_calc}\\
    \theta &= \theta - \alpha \nabla_{\theta} L(y,\hat{y}) \label{eq:loss_update}
\end{align}

We employ SHAP (SHapley Additive exPlanations) \cite{lundberg2017a} to additively interpret the contribution of each feature to the GBDT ensemble. This helps us in model selection to add and eliminate LMs and GNNs in a constrained practical application. SHAP calculates the features' contribution using a recursive algorithm that traverses the decision tree and aggregates the contributions of {every feature at each node}. The SHAP values for a feature are the average change in the prediction caused by that feature, averaged over all possible ways that the feature could be used in the decision tree. For GBDTs, the SHAP value of a feature $i$ is calculated as:
\begin{equation}
    \Omega_i = \sum_{F \subseteq \omega_X} \frac{|F|!(|X| - |F| - 1)!}{|X|!} (\Omega_{F \cup {i}} - \Omega_{F})
    \label{eq:shape}
\end{equation}
where $\Omega_i$ is the SHAP value for feature $i$, $F$ is a subset of features, $X$ is the set of all features, and $\omega_X$ is the set of all possible subsets of $X$.
The SHAP values enable us to rank a model's contribution and use it for model selection. Note that, due to the additive nature of SHAP values, we can compute a model's contribution without re-training all previous candidates. 
\subsection{Model Components}
\label{sec:model-pipelines}
Computational constraints significantly vary in the training and inference phase in industry settings. The training phase optimizes for model accuracy and generalizability over broader datasets, whereas the inference phase must additionally optimize for model latency under constrained computational resources\footnote{Complete details of the computational setup are presented in Section \ref{sec:implementation}.}. In this section, we explain our model's training phase, model selection setup, and inference phase for a practical environment.

\textbf{Training Phase.} During the training phase, we have access to a significant amount of both CPU and GPU memory. Thus, we can optimally parallelize our data processing and model training steps. Additionally, we note that e-commerce queries have a $20\%-25\%$ monthly update rate, i.e., we receive only $20\%-25\%$ new queries every month while the rest overlap with previously processed queries. Thus, given the monthly cycle of model update, the time-intensive operations of Graph Construction and De-noising Product Information do not significantly affect the model pipelines in practice. However, we need to ensure that the models are compliant with industry standards, and thus, we optimize our hyperparameters (provided in Section \ref{sec:implementation}) in a restricted search space (model size limited to 2.5 GB in our case). The training phase receives an input query-product pair with its corresponding relevance label. The input pair is processed using the graph signals and input to update the parameters of candidate LMs and GNNs through training. The final trained models are stored as checkpoints. {Algorithm \ref{alg:training} in Appendix \ref{app:algorithms} presents the model's training flow.}

\textbf{Model and Relation Selection.} In this phase, our goal is to restrict the number of candidate LMs and relation-based GNN models based on the inference constraints, while preserving the best performance. For an inference metric $\Lambda$ (additive units such as GPU VRAM, CPU RAM, or inference time) with constraint $K$ and the set of candidate models $\{LM\}_{i=1}^{|LM|}$ and $\{GNN\}_{j=1}^{|GNN|}$, this phase selects candidates $M\subseteq LM \cup GNN$, with the following constraints:
\begin{align}
    \sum_{m \in M}\Lambda_m \leq K;\quad \sum_{m \in M}\Omega_m\geq\sum_{f \in F-M}\Omega_f
\end{align}  We need to handle the selection in two scenarios: (i) initial setup where $M=\varnothing$ and (ii) continuous deployment where $M\neq\varnothing$. For the initial setup, we rank the candidate models according to their SHAP values (computed using Eq. (\ref{eq:shape})) and incrementally add them to $M$ till the constraints hold. In the case of continuous deployment, we follow a last-in, first-out (LIFO) strategy and compare the candidate model $f$ against the last-inserted model $m_l \in M$. We use the new candidate only if $\Omega_f > \Omega_{m_l}$ and $\sum_{m \in M-\{m_l\}+\{f\}}\Lambda_m \leq K$. {The model selection algorithm is provided in Appendix \ref{app:algorithms}.}

\textbf{Inference Phase.} In the inference phase, the model must handle a large volume of query-product pairs, and hence, the scalability and latency become crucial considerations. To address these concerns, we relegate the Graph Computation and De-noising Product Information modules to the pre-computation step. Thus, during inference, we load {\model} parameters on the available GPUs and process the query-product pairs in parallel batches (retrieving the graph and de-noised product information in constant time). One potential issue with this method is that pre-computed graph and de-noised product information may not be available for {infrequent} query-product pairs. However, in typical e-commerce settings, we have observed that this information is available for approximately 80\%-85\% of queries each month. For the remaining queries, we can still use a subset of candidate models (which do not rely on graph information) and obtain suboptimal results even with incomplete or noisy product information. It should be noted that, because the candidate models' predictions are independent, we can set up reliability-availability trade-offs\footnote{A focus on reliability would imply all models succeed, whereas, availability is possible if any model succeeds.} according to the downstream use case. {Our model's inference pipeline is presented in Appendix \ref{app:algorithms}.}

\subsection{Implementation Details}
\label{sec:implementation}
{\model} is implemented using the PyTorch framework with the huggingface library for the LM models and the PyG library for the graph neural networks. The model is trained on sixteen Nvidia V100 GPUs and optimized using AdamW \cite{loshchilov2018decoupled} with standard beta parameters of 0.9 and 0.999, weight decay rate of 0.01, and mini-batch gradient descent of batch size 64. The language models encode the input tokens with a maximum length of $|t|=512$ into a $d=768$-dimensional vector to be classified into $l=4$ labels. For GNNs, we use $|\psi|=5$ behavioral signals in a $K=2$-layer GNN network with $l=4$ final output labels. The graph extraction module obtains the local $k=2$-hop neighborhood with a maximum size of 100 neighbors. The GBDT ensemble is implemented using the LightGBM library \cite{ke2017lightgbm} with a cross-entropy objective, 1500 iterations, 0.005 learning rate, 15 leaves, 15 depth, and 200 bagging frequency. The inference setup is constrained to four V100 GPUs, with a time limit of 15 milliseconds per query-product sample. To avoid loading overheads, we use LuaJIT compiler to load {\model}'s parameters for inference. The implementation code of our model is available here\footnote{\url{https://github.com/amazon-science/graph-lm-ensemble}}. The dataset of query-product pairs is publicly available \cite{reddy2022shopping}.

\section{Experimental Results}
\label{sec:experimental}
Our experimental setting investigates the performance of our model for the ESCI classification problem and analyzes its components in practical environments. More specifically, we study the following research questions (\textbf{RQs)}:
\begin{itemize}
    \item[\textbf{RQ1.}] How does {\model}'s performance compare against current alternatives on the search relevance tasks of ESCI classification and irrelevant classification?
    \item[\textbf{RQ2.}] Is {\model} compatible with industry constraints? How does the performance differ in such scenarios? 
    \item[\textbf{RQ3.}] Is the ensemble method of feature aggregation better than other techniques?
    \item[\textbf{RQ4.}] How do we make model decisions using the additive SHAP values?
\end{itemize}
\subsection{Datasets and Baselines}
For our experiments on search relevance tasks, we utilize the publicly available ESCI dataset \cite{reddy2022shopping}. The dataset contains $\approx$2 million real-world query-product pairs with manually annotated ESCI labels, collected from the regions of the United States (US), Japan (JP), and Spain (ES). The regions are diverse in their languages, geographical location, and user behavior. This dataset is enriched with five additional behavioral signals: impressions (user viewed the product), clicks (user clicked on the product), adds (user added the product to their cart), purchases (user purchased the product), and consumes (user finally consumed the product). As shown in Figure \ref{fig:edge_corr}, we found that certain behavioral signals such as adds and purchases have a strong correlation with target ESCI labels, whereas others such as impressions and clicks have a denser availability. The dataset does not contain any customer information or personally identifiable information. From the perspective of an e-commerce search engine, we evaluate our model on two real-world tasks of ESCI classification (applied in complementary product recommendation \cite{hao2020pcompanion}) and irrelevant classification (used for improving user experience by increasing the precision of product recommendation). Each task leads to a unique imbalanced class distribution, which our model handles with class weights in the loss function. For the experiments, we hold out the test set (10\% of the dataset) for evaluation and perform a five-fold cross-validation with a train-validation split of 4:1. Further details on the datasets' edge distribution and edge correlation are provided in Table \ref{tab:edge_dist} and Figure \ref{fig:edge_corr}, respectively. {Note that densely available behavioral signals such as impressions and clicks have a low correlation to the target ESCI labels when compared to sparser signals such as clicks and adds. Therefore, in practical applications, it is crucial to carefully weigh the trade-off between reliability and availability when selecting which signals to use.} The dataset for the query-product pairs is available here\footnote{\url{https://github.com/amazon-science/esci-data}}.
\begin{table}[htbp!]
\vspace{-.6em}
    \centering
    \footnotesize
    \caption{Distribution of the edges in the dataset. The columns represent the number of edges and its proportion in the overall dataset.}
    \vspace{-.6em}
    \label{tab:edge_dist}
\vspace{-.6em}
\begin{tabular}{l|cc}
\hline
\textbf{Relation}    & \textbf{Number}     & \textbf{Proportion (\%)} \\\hline
\textbf{impressions} & 30,256,494 & 40.15      \\
\textbf{clicks}      & 38,739,273 & 49.86      \\
\textbf{adds}        & 3,886,923  & 5.25       \\
\textbf{purchases}   & 1,445,229  & 1.93       \\
\textbf{consumes}    & 118,440    & 0.17       \\
\textbf{exact}       & 1,451,542  & 1.72       \\
\textbf{substitute}  & 487,544    & 0.58       \\
\textbf{complement}  & 64,481     & 0.08       \\
\textbf{irrelevant}  & 223,804    & 0.27       \\\hline
\textbf{Total}       & 76,673,730 & 100        \\\hline  
\end{tabular}
\end{table}
\begin{figure}[htbp!]
    \centering
    \vspace{-1em}
    \includegraphics[width=.7\linewidth]{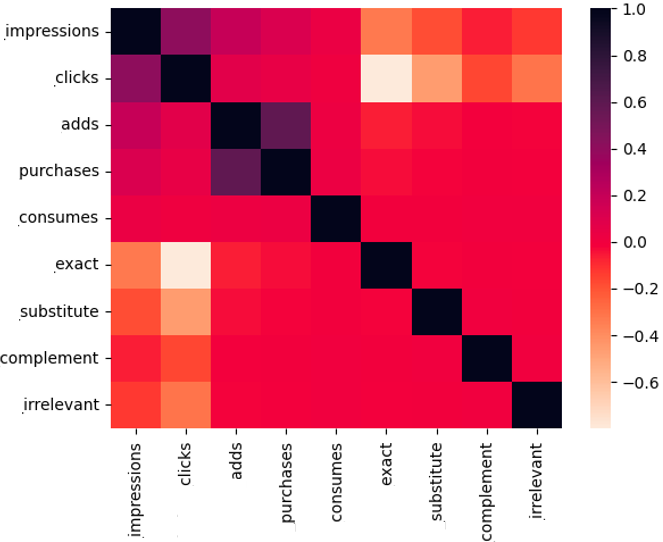}
    \vspace{-1em}
    \caption{Correlation between edges of the dataset. Note that dense signals such as impressions and clicks have a low correlation with target edges of exact, substitute, complement, and irrelevant, whereas sparser signals such as adds and purchases are highly correlated. Hence, we must carefully consider the reliability-availability trade-off in practice.}
    \label{fig:edge_corr}
    \vspace{-1em}
\end{figure}

For the baselines, we select (i) the popular language models of DeBERTa, COCOLM, BigBird, and M-DeBERTa, (ii) GNN models of GCN, GAT, and GraphSage, (iii) Relation-GNN models of GraphSage model trained using behavioral signal-attributed edges\footnote{the basic relations are impressions, adds, clicks, purchases, and consumes. However, we also use a homogeneous (All) and heterogeneous (Het-All) version of the graph with all the basic relations.}, and (iv) SALAM \cite{choudhary2022graph}, which is a graph-based language model for e-commerce engines. Language models only utilize the query-product text information. For graph models, the node features are initialized using semantic embeddings from LM models (M-DeBERTa). SALAM utilizes both the text information as well as the behavioral signals in its framework.
\begin{table*}[htbp!]
\caption{Performance comparison of {\model} against baselines for search relevance tasks of ESCI classification and irrelevant classification in the e-commerce regions of US, ES, and JP. Evaluation metrics (presented in the columns) include Accuracy (Acc), Macro-F1 (MacF1), and Weighted-F1 (WtF1). The best results for each metric are in bold font. All the improvements of {\model} over SALAM and best-performing models are statistically significant with a p-value threshold of 0.05.}
\label{tab:search-relevance}
\setlength{\tabcolsep}{2.5pt}
\vspace{-.6em}
\footnotesize
\begin{tabularx}{\textwidth}{ll|ccc|ccc|ccc|ccc|ccc|ccc}
         \hline
         \textbf{Task}&         & \multicolumn{9}{c|}{\textbf{ESCI Classification}} & \multicolumn{9}{c}{\textbf{Irrelevant Classification}}\\\hline
         \textbf{Model}& \textbf{Locale} & \multicolumn{3}{c|}{\textbf{United States (US)}} & \multicolumn{3}{c|}{\textbf{Spain (ES)}} & \multicolumn{3}{c|}{\textbf{Japan (JP)}} & \multicolumn{3}{c|}{\textbf{United States (US)}} & \multicolumn{3}{c|}{\textbf{Spain (ES)}} & \multicolumn{3}{c}{\textbf{Japan (JP)}} \\
         \textbf{Type}& \textbf{Model} & \textbf{Acc}   & \textbf{MacF1} & \textbf{WtF1}  & \textbf{Acc}   & \textbf{MacF1} & \textbf{WtF1}  & \textbf{Acc}   & \textbf{MacF1} & \textbf{WtF1} & \textbf{Acc}   & \textbf{MacF1} & \textbf{WtF1}  & \textbf{Acc}   & \textbf{MacF1} & \textbf{WtF1}  & \textbf{Acc}   & \textbf{MacF1} & \textbf{WtF1} \\\hline
\textbf{LM}       & \textbf{(M)DeBERTa}  & 84.42 & 70.00 & 84.42 & 76.84 & 66.99 & 76.84 & 75.91 & 64.33 & 75.91 & 95.71 & 87.10 & 95.61 & 86.52 & 46.39 & 80.27 & 90.20 & 81.52 & 90.18 \\
         & \textbf{COCOLM}      & 81.19 & 64.19 & 81.19 & -     & -     & -     & -     & -     & - & 95.18 & 84.58 & 94.91 & -     & -     & -     & -     & -     & -     \\
         & \textbf{BigBird}     & 86.58 & 74.40 & 86.58 & -     & -     & -     & -     & -     & - & 95.92 & 90.96 & 95.89 & -     & -     & -     & -     & -     & -      \\\hline
\textbf{GNN}      & \textbf{GraphSage}   & 75.29 & 46.91 & 71.85 & 63.79 & 40.61 & 59.23 & 64.67 & 40.93 & 60.84 & 90.41 & 47.49 & 85.87 & 86.52 & 46.41 & 80.27 & 86.22 & 46.45 & 79.86\\
         & \textbf{GCN}         & 71.47 & 34.53 & 64.15 & 60.34 & 34.78 & 55.46 & 60.99 & 35.25 & 55.28  & 90.40 & 47.49 & 85.85 & 86.54 & 46.39 & 80.28 & 86.20 & 46.42 & 79.82\\
         & \textbf{GAT}         & 72.14 & 38.99 & 66.75 & 60.65 & 35.60 & 56.13 & 61.92 & 37.23 & 57.22 & 90.43 & 47.49 & 85.86 & 86.53 & 46.39 & 80.27 & 86.22 & 46.44 & 79.85\\\hline
\textbf{Relation} & \textbf{Impressions} & 72.38 & 37.80 & 66.06 & 61.49 & 37.07 & 56.76 & 62.56 & 36.95 & 57.12 & 90.40 & 47.51 & 85.84 & 86.53 & 46.40 & 80.30 & 86.21 & 46.41 & 79.84\\
\textbf{GNN}      & \textbf{Adds}        & 73.56 & 41.95 & 68.81 & 61.62 & 37.57 & 57.51 & 62.79 & 39.15 & 58.89 & 90.41 & 47.48 & 85.87 & 86.53 & 46.40 & 80.30 & 86.20 & 46.46 & 79.85\\
         & \textbf{Clicks}      & 72.43 & 38.16 & 66.19 & 61.37 & 37.15 & 56.78 & 62.51 & 36.70 & 56.90  & 90.42 & 47.48 & 85.85 & 86.53 & 46.40 & 80.30 & 86.21 & 46.42 & 79.83\\
         & \textbf{Purchases}   & 73.92 & 43.27 & 69.76 & 61.80 & 37.19 & 57.34 & 62.84 & 39.22 & 58.93 & 90.43 & 47.49 & 85.85 & 86.53 & 46.41 & 80.29 & 86.22 & 46.44 & 79.86\\
         & \textbf{Consumes}    & 73.90 & 44.81 & 70.72 & 61.88 & 36.75 & 57.21 & 62.21 & 39.75 & 59.26 & 90.43 & 47.50 & 85.86 & 86.53 & 46.40 & 80.29 & 86.20 & 46.51 & 79.87\\
         & \textbf{All}         & 72.22 & 37.23 & 65.59 & 61.45 & 36.91 & 56.64 & 62.37 & 36.26 & 56.54  & 90.43 & 47.48 & 85.85 & 86.53 & 46.41 & 80.28 & 86.21 & 46.39 & 79.83\\
         & \textbf{Het-All}     & 72.35 & 37.80 & 65.94 & 61.52 & 36.35 & 56.33 & 62.43 & 36.59 & 56.83  & 90.42 & 47.49 & 85.84 & 86.55 & 46.40 & 80.28 & 86.21 & 46.44 & 79.85\\\hline
\textbf{GNN}   & \textbf{SALAM} & 83.82 & 67.17 & 84.61 & 75.77 & 60.72 & 76.49 & 75.74 & 63.71 & 74.51 & 89.82 & 73.17 & 90.61 & 81.77 & 44.72 & 78.49 & 86.74 & 70.71 & 84.51 \\
\textbf{+LM}   & \textbf{{\model} (Ours)} & \textbf{90.45} & \textbf{82.36} & \textbf{90.79} & \textbf{80.32} & \textbf{69.81} & \textbf{80.63} & \textbf{79.23} & \textbf{67.68} & \textbf{79.61} & \textbf{96.69} & \textbf{90.99} & \textbf{96.78} & \textbf{87.90} & \textbf{46.78} & \textbf{82.24} & \textbf{90.41} & \textbf{82.08} & \textbf{90.92}\\\hline
\end{tabularx}
\vspace{-.6em}
\end{table*}

\subsection{RQ1: Search Relevance}
In this experiment, we investigate the performance of our model relative to other baselines on the search relevance tasks of ESCI classification and irrelevant classification. The {\model} model predicts a single label for each query-product pair and the performance is evaluated using ground-truth labels on the metrics of accuracy, macro-F1, and {weighted F1 scores.\footnote{weights are given by the number of actual occurrences of the class in the dataset.}} We evaluate both the macro-F1 and weighted F1 scores due to class imbalance, and certain downstream tasks such as recommendation systems require the weighted F1 to be high. For other tasks, such as irrelevant detection, it will be ideal to obtain a high macro-F1.

From our experimental results, reported in Table \ref{tab:search-relevance}, we observe that our model consistently outperforms the baselines in both the tasks of ESCI classification and irrelevant classification. Among the baselines, we note that LM models tend to outperform the graph models, which underscores the importance of semantic features. Even in the case of LM models for the US region, we notice that the newer LM model (i.e., BigBird) outperforms the relatively older models such as DeBERTa and COCOLM, which strengthens our case for plug and play ensemble models with replaceable components. {\model}, with its ensemble framework, improves the performance of both standalone LMs and GNNs by 4\%-11\% and 20\%-75\%, respectively. Moreover, the model can also leverage further research advances within LM models and GNNs in a non-intrusive manner. This enables us to take advantage of the latest works in an efficient pipeline. Additionally, we also see that {\model} outperforms the SALAM model, which also utilizes both LM and GNN models. This is because SALAM gives more importance to graph features, which are not available during the evaluation phase. Thus, we conclude that an ensemble that can adapt its dependence on language and graph features according to the dataset is the best fit for a real-world industry scenario of search relevance.

\begin{table}[htbp!]
\vspace{-.6em}
\caption{Comparison of memory and processing requirements for different models in the training and inference pipelines. The Columns indicate the number of parameters (Param), pre-training time (PTT), and fine-tuning time (FTT) in seconds per epoch, inference time (IT) in milliseconds per sample, VRAM requirement (Mem), and disk space requirement (Disk) in Gigabytes.}
\label{tab:practical}
\vspace{-.6em}
\setlength{\tabcolsep}{3pt}
\footnotesize
\begin{tabular}{l|cccccc}
\hline
\textbf{Model}        & \textbf{Param} & \textbf{PTT}  & \textbf{FTT} & \textbf{IT}    & \textbf{Mem}  & \textbf{Disk}  \\\hline
\textbf{LM}           & 230M  & 128K & 215 & 10.67 & 1.01 & 16.2  \\
\textbf{GNN}          & 70M   & 133K & 169 & 9.87  & 0.49 & 39.8  \\
\textbf{Relation-GNN} & 150M  & 130K & 217 & 10.83 & 1.38 & 22.1  \\
\textbf{SALAM}        & 279M  & 135K & 231 & 11.05 & 2.57 & 169.1 \\
\textbf{{\model} (Ours)} & 450M  & 138K & 220 & 10.67 & 2.38 & 160.1 \\
\textbf{{\model} (Red)} & 290M & - & - & 10.50 & 0.98 & 160.1\\\hline
\end{tabular}
\vspace{-.4em}
\end{table}
\subsection{RQ2: Practical Environment}
To comprehend the suitability of {\model} for industry settings, we compare its memory and processing requirements against the LM- and GNN-based baselines. We evaluate the models on the basis of their number of parameters (Param), pre-training time (PTT), fine-tuning time (FTT), inference time (IT), GPU RAM requirement (Mem), and disk space requirement (Disk).

Results shown in Table \ref{tab:practical} demonstrate that our model's computational and memory requirements are comparable to that of the available alternatives and differ only by a negligible margin of 3\%-5\%. This margin is easily manageable due to the monthly update rate of query-product pairs, as discussed in Section \ref{sec:model-pipelines}. An important challenge, however, is the additional time taken by the Graph Extraction and De-noising product information modules. In our pipeline, we shift these modules to the pre-computation step and save the results as hash tables to enable constant-order retrieval. However, this leads to a lack of availability of certain graphs for the query-product pairs in the evaluation set and also adds to the disk space requirement, which increases proportionally to the number of unique query-product pairs. For the problem of search relevance, our model can inductively handle the graph unavailability problem by returning the results from the language model. Furthermore, the additional disk requirement is not a concern since it is relatively inexpensive. However, further applications of our model should consider these limitations.
\begin{table*}[!htbp]
\setlength{\tabcolsep}{3pt}
\footnotesize
\caption{Comparison of different feature aggregation methods (MLP and attention) and the constrained model variant ({\model} (Red)) of our model for the search relevance tasks of ESCI classification and irrelevant classification. Evaluation metrics presented in the columns are accuracy (Acc), macro-F1 (MacF1), and weighted F1 (WtF1). The best results are in bold font.}
\label{tab:feature-aggregation}
\vspace{-.6em}
\begin{tabular}{l|ccc|ccc|ccc|ccc|ccc|ccc}
\hline
\textbf{Task} & \multicolumn{9}{c|}{\textbf{ESCI Classification}} & \multicolumn{9}{c}{\textbf{Irrelevant Classification}}\\\hline
\textbf{Locale} & \multicolumn{3}{c|}{\textbf{United States (US)}} & \multicolumn{3}{c|}{\textbf{Spain (ES)}} & \multicolumn{3}{c|}{\textbf{Japan (JP)}} & \multicolumn{3}{c|}{\textbf{United States (US)}} & \multicolumn{3}{c|}{\textbf{Spain (ES)}} & \multicolumn{3}{c}{\textbf{Japan (JP)}} \\\hline
\textbf{Model} & \textbf{Acc}  & \textbf{MacF1} & \textbf{WtF1}  & \textbf{Acc}    & \textbf{MacF1} & \textbf{WtF1}  & \textbf{Acc} & \textbf{MacF1} & \textbf{WtF1}& \textbf{Acc} & \textbf{MacF1} & \textbf{WtF1}  & \textbf{Acc} & \textbf{MacF1} & \textbf{WtF1}  & \textbf{Acc} & \textbf{MacF1} & \textbf{WtF1}  \\\hline
\textbf{MLP} & 81.18  & 74.06 & 79.92 & 69.64  & 60.76 & 71.94 & 70.41  & 56.98 & 69.15 & 86.78 & 81.82 & 85.19 & 76.21 & 40.72 & 73.38 & 80.35 & 69.10 & 78.97\\
\textbf{Attention} & 85.20 & 76.77 & 85.46 & 75.22  & 64.05 & 75.18 & 74.04  & 62.09 & 74.47 & 91.08 & 84.81 & 91.10 & 82.32 & 42.92 & 76.68 & 84.49 & 75.30 & 85.05\\\hline
\textbf{{\model} (Ours)} & \textbf{90.45}  & \textbf{82.36} & \textbf{90.79} & \textbf{80.32}  & \textbf{69.81} & \textbf{80.63} & \textbf{79.23}  & \textbf{67.68} & \textbf{79.61} & \textbf{96.69} & \textbf{90.99} & \textbf{96.78} & \textbf{87.90} & \textbf{46.78} & \textbf{82.24} & \textbf{90.41} & \textbf{82.08} & \textbf{90.92}\\
\textbf{{\model} (Red)} & 89.04 & 80.86 & 89.31 & 79.03 & 68.43 & 79.50 & 78.07 & 66.28 & 78.36 & 95.18 & 89.33 & 95.20 & 86.49 & 45.86 & 81.09 & 89.09 & 80.38 & 89.49 \\\hline
\end{tabular}
\vspace{-.6em}
\end{table*}
\subsection{RQ3: Feature Aggregation}
To confirm the effectiveness of the GBDT ensemble as a feature aggregation mechanism, we compare its performance to other commonly used mechanisms of using multi-layer {perceptrons} (MLPs) layer and attention layers. Both the MLP and attention layer perform their feature aggregation in a latent space. This leads to a lack of interpretability and generalization over a new data distribution. In addition, both these approaches will require an expensive re-training step for the entire model. In contrast, our GBDT based ensemble consists of decision trees that are interpretable using additive SHAP values. Also, they are trained on the outputs obtained from different models, and thus, the training of the {aggregation mechanism (ensemble)} for a new data distribution does not require the re-training of {component graph and language models}. Furthermore, from Table \ref{tab:feature-aggregation}, we note that GBDT ensembles perform better by 6\%-9\% at feature aggregation than their alternatives because they can better generalize over the evaluation dataset.

\subsection{RQ4: Additive Explanations}
In this experiment, we study the interpretability of our model by analyzing the contribution of LM and GNN models and different behavioral signals using additive SHAP explanations and aim to compress the model for inference by removing the models with lower contribution.
\begin{figure}[htbp!]
\centering
\vspace{-.6em}
\begin{subfigure}{.49\linewidth}
  \centering
  \includegraphics[width=\linewidth]{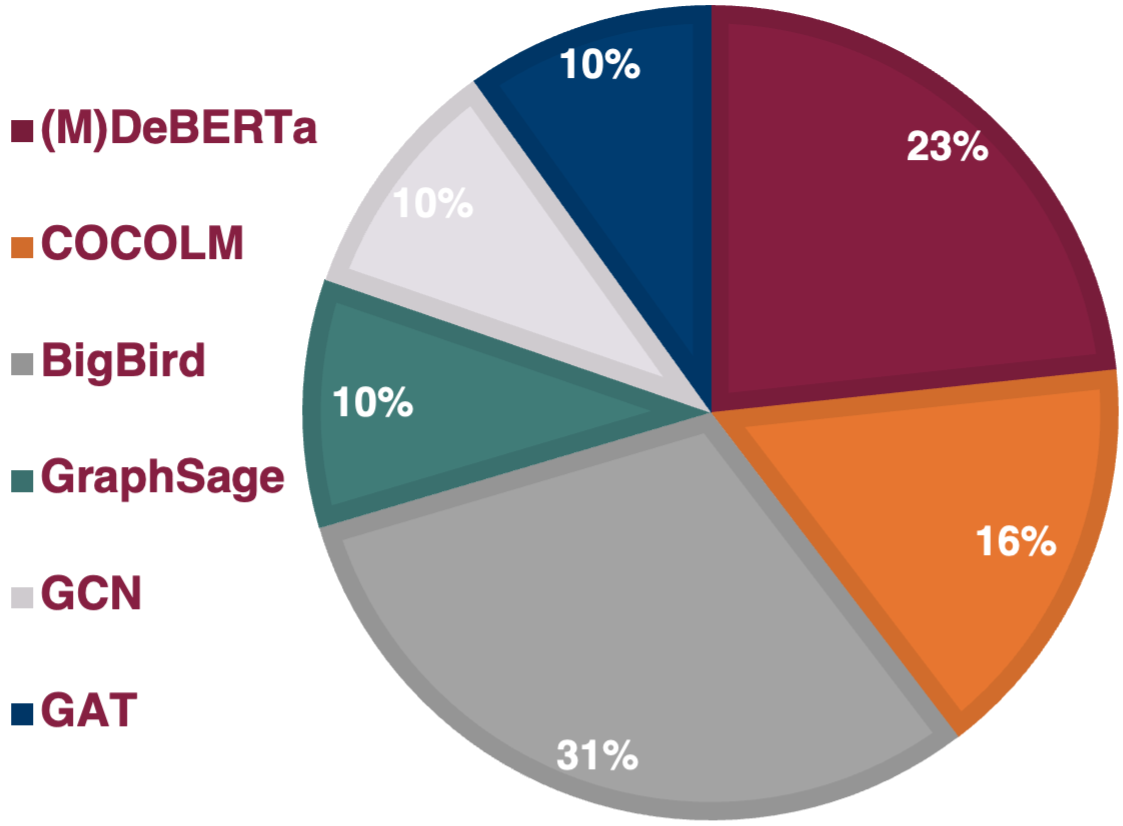}
  \caption{Model contribution}
  \label{fig:model-contribution}
\end{subfigure}%
\begin{subfigure}{.49\linewidth}
  \centering
  \includegraphics[width=\linewidth]{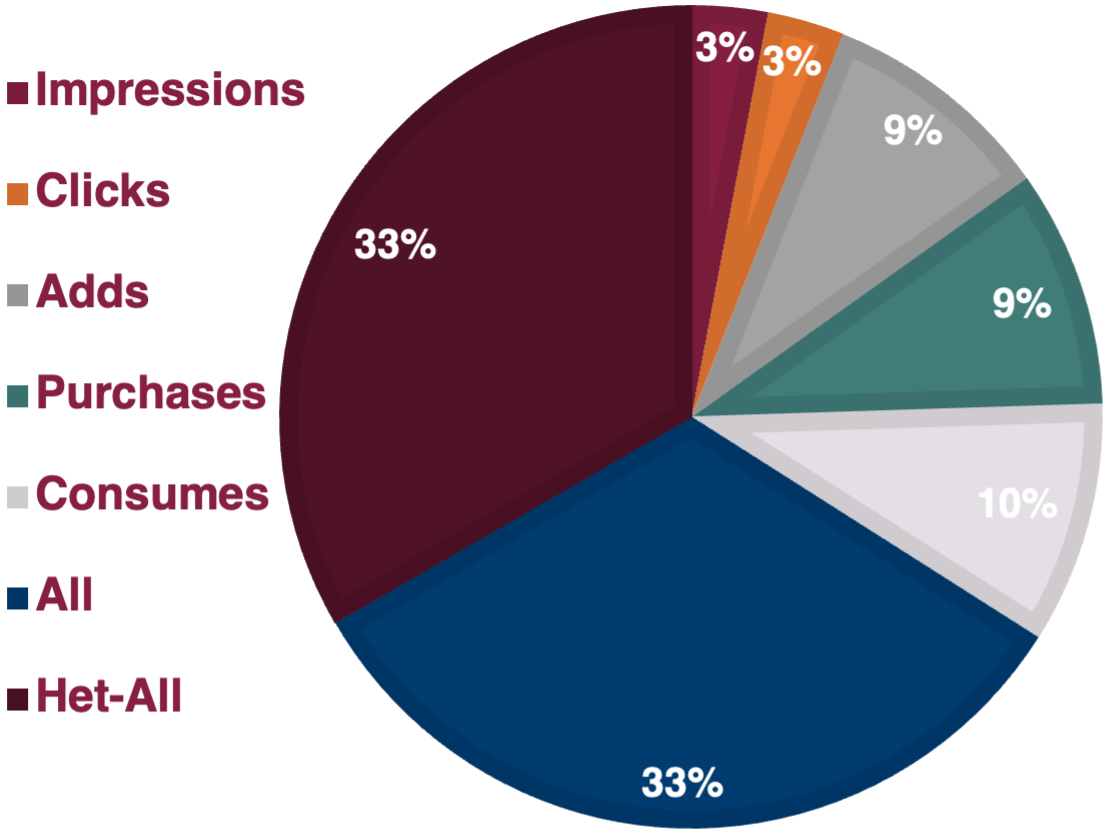}
  \caption{Relation contribution}
  \label{fig:relation-contribution}
\end{subfigure}
\vspace{-.6em}
\caption{SHAP value-based additive contribution of different features to the overall performance of our model. (a) provides the contribution of different models, and (b) shows the contribution of different relations.}
\vspace{-.6em}
\label{fig:shap-contributions}
\end{figure}

From Figure \ref{fig:model-contribution}, we observe that LM models (around 16\%-31\%) contribute significantly higher to our model's performance compared to GNN models (at 10\% each). Diving deeper into the contribution of different relations, we notice that GNN models that use all features either as homogeneous graphs or heterogeneous graphs have the highest contribution at 33\%, followed by strong behavioral signals such as adds and purchases at 9\%-10\%. Finally, relatively weak behavioral signals have an insignificant contribution of 3\% to the model performance. Based on these contributions, we construct an ensemble of reduced size, {\model} (Red), with the highest contribution models of (M)DeBERTa, BigBird, and GraphSage with Het-All relations. We observe that the reduced ensemble leads to a $57.8\%$ reduction in the number of parameters (Table \ref{tab:practical}) with a performance reduction of only $1.6\%$ (Table \ref{tab:feature-aggregation}).

\section{Discussion}
\label{sec:discussion}

This section describes a strategy to deploy {\model} in an industrial setting and then understand its impact on the search engine.

\textbf{Deployment Strategy:} As explained in Section \ref{sec:model-pipelines}, {\model} can be trained and ensembled offline using annotated labels. The trained models can then be stored as checkpoints for inference. These checkpoint models will be loaded in parallel GPUs using LuaJIT platform, which enables faster real-time inference by removing the overhead of loading weights for each iteration. During inference, the model processes query-product pairs in batches and saves the predicted labels for further utilization in downstream tasks. The integration of the {\model} method into existing workflows that utilize query-product pairs as input for classification is a simple task with minimal modifications required. The graph preprocessing step is also implemented with efficient caching and real-time lookup mechanism {with constant order retrieval}. In particular, given the long-tailed distribution of product search queries, we can pre-compute query and product neighborhoods for the most frequently seen samples. Additionally, {efficient online inference methods \cite{jiang2018efficient}} are employed to compute representations for less frequently seen queries in real time. The inclusion of graph information results in significant performance gains, while the impact on inference time is negligible. Thus, the deployment of {\model} is a feasible and viable investment for practical applications.

\textbf{Impact on Search Engine:} In this section, we present the impact of incorporating an ESCI classifier in a product search engine. The ranking of e-commerce products relies on lexical, behavioral, and semantic matching. However, behavioral data can be unreliable, leading to biases in the data used to train these models. By implementing an ESCI-based classification at the final stage, or using these {classification predictions} as inputs for downstream models, these issues can be mitigated. Additionally, the results from these models can be utilized for customer messaging, providing explanations for why a particular item was shown in response to a query. Improving the ESCI classifiers' performance can both enhance the quality of product search and the customers' trust in these systems, ultimately allowing practitioners to better serve their end-users.

\section{Conclusion}
\label{sec:conclusion}
In this paper, we presented PP-GLAM, a modular ensemble of LMs and relational GNNs that utilizes GBDTs for flexible model selection in a practical environment. Through our experimental evaluation on a multi-regional public e-commerce dataset, we have shown the effectiveness of PP-GLAM on search relevance and irrelevant detection tasks of ESCI classification, outperforming current alternatives. We have also demonstrated the compatibility of PP-GLAM with industry settings and highlighted the benefits of using ensemble methods as an interpretable strategy to aggregate semantic and behavioral signals and efficiently select the most impactful models. Additionally, we have detailed a deployment strategy for integrating our framework in practical settings with dynamic data sources. Overall, our approach presents a promising step towards improved e-commerce product search.

\balance
\newpage
\bibliographystyle{ACM-Reference-Format}
\bibliography{sample-base}

\appendix
\section{Algorithms}
\label{app:algorithms}

Algorithms \ref{alg:selection} and \ref{alg:inference} provide the algorithm for our model's selection pipeline, and our model's inference pipeline, respectively. Algorithm \ref{alg:training} presents the model's training flow. 

\removelatexerror
\begin{algorithm}[H]
    \SetAlgoLined
    \KwIn{Current model set $M$, candidate models $LM$, $GNN$, inference metric $\Lambda$, constraint $K$;}
    \KwOut{New model set $M$;}
    \uIf{$M=\varnothing$}{
    {\color{blue}{ \# Sort candidate models based on SHAP values}}\\
    $R = sort_{\Omega_f}\left(\{LM \cup GNN\}\right)$\\
    Initialize $i=1$, $\Lambda = 0$;\\
    }
    \Else{
    {\color{blue}{ \# Sort candidate and $|LM \cup GNN|$ least performing models in $M$}}\\
    $lst(M) = \argmin_{\Omega,|LM ~\cup~ GNN|}(M)$\\
    $R = sort_{\Omega_f}\left(\{LM \cup GNN \cup lst(M)\}\right)$\\
    Initialize $i=1$, $\Lambda = \sum_{m\in M \setminus lst(M)} \Lambda_m$;\\
    }
    \While{$\Lambda \leq K$ \& $i \leq |R|$}{
        $\Lambda = \Lambda + \Lambda_m;$\\
        $M = M \cup R[i];$\\
    }
    \Return{$M$}
    \caption{Model Selection.}
    \label{alg:selection}
\end{algorithm}
\begin{algorithm}[H]
	\SetAlgoLined
	\KwIn{Query-product pairs $(Q,P)=\{(q,p)\}$, Pre-computed neighborhoods $\mathcal{G}^\psi=\{\mathcal{G}^\psi(q,p)\}$, Ground truth $y$;}
	\KwOut{Predictor $P_\theta, \theta$;}
	Initialize model parameters $\theta$;\\
	\For{number of epochs; until convergence}
	{
	Initialize loss $l=0$;\\
	\For{$(q,p) \in (Q,P), \mathcal{G}^\psi(q,p) \in \mathcal{G}^\psi$}
	{
            Tokenize input $t=q\|[SEP]\|p'$;\\
	    {\color{blue}{ \# Process through language models}\\}
            \For{$LM_i \in LM$}
            {
                $\hat{y}_i^{LM}(t) = \phi_d^l(LM_i(t))$; via Eq. (\ref{eq:lm})\\
            }
            $\hat{Y}_{LM}(t) = \left\{\hat{y}^{LM}_i(t)\middle| i \in [1,|LM|]\right\}$; via Eq. (\ref{eq:lm_agg})\\
            {\color{blue}{ \# Aggregate node neighborhoods}\\}
            \For{$GNN_i \in GNN$}
            {
                $h_{k+1} = \sigma\left(D^{\zeta-1}\mathcal{G}^{\psi}(q,p)D^{\zeta}h_kW_k\right)$\\
                $\hat{y}_i^{GNN}(h_0) = \phi_d^l(h_{k+1}[i_q]\|h_{k+1}[i_p])$; via Eq. (\ref{eq:gnn})\\
            }    
            $\hat{Y}_{GNN}(h_{0}) = \left\{\hat{y}^{GNN}_i(h_{0})\middle| i \in [1,|GNN|]\right\}$; via Eq. (\ref{eq:gnn_agg})\\
            {\color{blue}{ \# GBDT Ensemble}\\}
            $\hat{Y}=\hat{Y}_{LM}(t)\|\hat{Y}_{GNN}(h_{0})\|f$\\
            $\hat{y} = \frac{1}{T}\sum_{t=1}^T o_t(\hat{Y})$; via Eq. (\ref{eq:gbdt})\\
            {\color{blue}{ \# Loss Calculation }\\}
            $L(y,\hat{y}) = -\sum_{y \in Y}\left(ylog(\hat{y})+(1-y)log(1-\hat{y})\right)$; via Eq. (\ref{eq:loss_calc})\\
            $l = l + L(y,\hat{y})$;\\
      }
      {\color{blue}{ \# Update parameters}}\\
      $\theta = \theta - \alpha \nabla_{\theta} L(y,\hat{y})$; via Eq. (\ref{eq:loss_update})\\
      }
      \Return{$P_\theta, \theta$}
      \caption{{\model} training flow.}
      \label{alg:training}
\end{algorithm}

\begin{algorithm}[H]
	\SetAlgoLined
	\KwIn{$\{(q,p)\}$, $\{\mathcal{G}^\psi(q,p)\}$, Model set $M$;}
	\KwOut{Label $\hat{y}$;}
        $t=q\|[SEP]\|p'$;\\
        \For{$LM_i \in LM_M$}
        {
            $\hat{y}_i^{LM}(t) = \phi_d^l(LM_i(t))$; via Eq. (\ref{eq:lm})\\
        }
        $\hat{Y}_{LM}(t) = \left\{\hat{y}^{LM}_i(t)\middle| i \in [1,|LM|]\right\}$; via Eq. (\ref{eq:lm_agg})\\
        \For{$GNN_i \in GNN_M$}
        {
            $h_{k+1} = \sigma\left(D^{\zeta-1}\mathcal{G}^{\psi}(q,p)D^{\zeta}h_kW_k\right)$\\
            $\hat{y}_i^{GNN}(h_0) = \phi_d^l(h_{k+1}[i_q]\|h_{k+1}[i_p])$; via Eq. (\ref{eq:gnn})\\
        }    
        $\hat{Y}_{GNN}(h_{0}) = \left\{\hat{y}^{GNN}_i(h_{0})\middle| i \in [1,|GNN|]\right\}$; via Eq. (\ref{eq:gnn_agg})\\
            $\hat{Y}=\hat{Y}_{LM}(t)\|\hat{Y}_{GNN}(h_{0})\|f$\\
            $\hat{y} = \frac{1}{T}\sum_{t=1}^T o_t(\hat{Y})$; via Eq. (\ref{eq:gbdt})\\
        \Return{$\hat{y}$}
	\caption{{\model} inference flow.}
	\label{alg:inference}
\end{algorithm}
\end{document}